\documentclass[aps,pre,twocolumn,groupedaddress]{revtex4}
\usepackage{graphicx}
\usepackage{dcolumn}
\usepackage{bm}
\usepackage[caption=false]{subfig}
\usepackage{placeins}

\begin{document}
\title{A multi-body finite element model for hydrogel packings: Linear response to shear}

\author{Ahmed Elgailani}
\email{elgailani.a@northeastern.edu}

\author{Craig E Maloney}
\email{c.maloney@northeastern.edu}
\affiliation{Department of Mechanical and Industrial Engineering, Northeastern University}

\date{\today}

\begin{abstract}
We study a multi-body finite element model of a packing of hydrogel particles using the Flory-Rehner constitutive law to model the deformation of the swollen polymer network. 
We show that while the dependence of the pressure, $\Pi$, on the effective volume fraction, $\phi$, is virtually identical to a monolithic Flory material, the shear modulus, $\mu$, behaves in a non-trivial way.
$\mu$ increases monotonically with $\Pi$  from zero and remains below about $80\%$ of the monolithic Flory value at the largest $\Pi$ we study here.
The local shear strain in the particles has a large spatial variation.
Local strains near the centers of the particles are all roughly equal to the applied shear strain, but the local strains near the contact facets are much smaller and depend on the orientation of the facet. 
We show that the slip between particles at the facets depends strongly on the orientation of the facet and is, on average, proportional to the component of the applied shear strain resolved onto the facet orientation. 
This slip screens the stress transmission and results in a reduction of the shear modulus relative to what one would obtain if the particles were welded together at the facet.
Surprisingly, given the reduction in the shear modulus arising from the facet slip, and the spatial variations in the local shear strain inside the particles themselves, the deformation of the particle centroids is rather homogeneous with the strains of the Delaunay triangles having fluctuations of only order $\pm 5\%$.
These results should open the way to construction of quantitative estimates of the shear modulus in highly compressed packings via mean-field, effective-medium type approaches.  
\end{abstract}


\maketitle
\section{Introduction}
Particles made of hydrogels --  chemically cross-linked polymeric materials which swell when they absorb water -- are important in industries such as pharmaceuticals, bioengineering, agriculture, food science, and cosmetics~\cite{doi:https://doi.org/10.1002/9783527632992.ch17,doi:https://doi.org/10.1002/9783527632992.ch14,doi:https://doi.org/10.1002/9783527632992.ch16,doi:https://doi.org/10.1002/9783527632992.ch15}.
Owing to their softness and the in situ tunability of their size via controlled swelling, they have also been used more recently as a test bed for basic ideas in the physics of jamming and the glass transition~\cite{ISI:000289354500003,Mattsson:2009aa,ISI:000334494000013,ISI:000276199300058,ISI:000181015900058,ISI:000165556600044,ISI:000183582800003}.
The basic statistical mechanical description of the mechanical properties of macroscopic, monolithic pieces of hydrogel material goes back to Flory and Rehner~\cite{doi:10.1063/1.1723791,doi:10.1063/1.1723792,doiBook,colbyBook}, however, the properties of suspensions or packings of \emph{particles} made out of the hydrogel are much more subtle and have been the subject of active research only over the past couple of decades.  

Typically, in experiments, the particles in a suspension are initially allowed to freely swell at low number density, then,
once fully swollen, are subjected to osmotic confinement with a membrane~\cite{ISI:000183582800003,ISI:000297561400020,ISI:000297794700001} or to centrifugation~\cite{ISI:000282681400003}, forcing the swollen particles into persistent contact and making a jammed solid where the particles are no longer free to diffuse and pass each other without incurring an energetic penalty associated with deformation of the polymer network.
At low degrees of confinement, the particles remain essentially spherical, and contacts between them are only slightly deformed.
The mechanics and energetics of the interactions in this regime should be reasonably well described by standard Hertzian contact mechanics~\cite{Johnson:1985aa} treating one contact at a time resulting in a pairwise description of the energy of the whole system. 
However, as the confinement increases, the inter-particle contacts will develop into curved facets~\cite{Scheffold:2020aa,doi:10.1063/1.5026100}, and, at large enough confinement, the void space with pure solvent will completely disappear altogether.
In foams and emulsions at such large particle volume fractions, it is known that not only does linear contact mechanics fail to describe the interactions at a given contact, but even worse, pairwise descriptions of any kind become qualitatively inaccurate in this regime~\cite{ISI:000396024800007,WEAIRE2017491,HOHLER201919}, and there is no reason to think that the micro gel packings of interest here will be any different.
One must consider in detail the deformation inside the particles.
It is this strongly confined, fully-faceted, regime which is of primary interest to us here.

Several groups~\cite{ISI:000183582800003,ISI:000297561400020,ISI:000297794700001} have measured the elastic modulus, $G$, of highly compressed microgel packings under osmotic confining pressure, $\Pi$, at increasing nominal volume fraction, $\phi$.
In all cases, one observes a rapid increase of $\Pi$ and $G$ from, essentially, zero as the particles are first forced into contact at the jamming transition when $\phi$ reaches the random close packing point $\phi_J$.
The observed behavior of $\Pi$ and $G$ is qualitatively consistent with the jamming picture~\cite{ISI:000184582400020} where $\Pi$ and $G$ are both zero below the jamming point, $\phi_J$, and, near jamming, scale as non-trivial power of $\delta\phi=\phi-\phi_J$. 
Precise scaling laws for $\Pi$ and $G$ are difficult to obtain in experiments, but in all experiments, there is an obvious and dramatic onset of both $\Pi$ and $G$ near a critical $\phi$ value. 
At higher $\phi$, away from jamming on approach to the fully faceted regime, the situation becomes more complicated.
Cloitre et. al.~\cite{ISI:000183582800003}, have shown that there is a transition from a more pronounced to a less pronounced dependence of $G$ on $\phi$ and argued that it occurs at the point at which the assembly becomes fully faceted with no void regions of pure solvent remaining.
Liétor-Santos et. al. ~\cite{ISI:000297794700001} showed that $G$, $\Pi$, and, the compression modulus, $K$, all scaled with $K_p$, the compression modulus of an isolated particle osmotically confined to the same average particle size as in the packing, so that $G/K_p$, $K/K_p$, and $\Pi/K_p$ were all universal constants in the compressed packing independent of the degree of compression of the packing or the degree of crosslinking of the particles.
Menut et. al.~\cite{ISI:000297561400020} have studied a variety of particles with different cross-linking densities and sizes.
They argued that sufficiently far above the jamming point, the density dependence of the modulus followed the trend that would be expected for a monolithic Flory material~\cite{doi:10.1063/1.1723791,doi:10.1063/1.1723792}.
Here, we find, what is perhaps the simplest non-trivial outcome one could have expected:  a pressure vs. $\phi$ behavior almost precisely the same as the monolithic Flory material and a universal shear modulus vs. pressure curve for systems with different cross-linking density when the modulus and pressure are both scaled by the Flory pressure, $NkT$, where $N$ is the density of cross-links in the dry reference state.


\section{Model}
Previous modeling work has proceeded along many fronts.
Many groups assume pair-wise additive interactions between particles even in a very high-density regime far from jamming~\cite{Bergman:2018aa,ISI:000408781200082,Scheffold:2020aa,Conley:2019aa,ISI:000286746500001}, sometimes modified to attempt to account for $\phi$ dependent effective interactions ~\cite{ISI:000388210400014,C8SM01153B,Higler:2018aa,PhysRevLett.76.3448,PhysRevE.56.3150,HOHLER201919,C6SM01567K}, which are, however, still \emph{pair-wise} contact interactions.
As we mention above, this approach is known to fail in foams and emulsions due to strong many-body effects in the particle-particle forces~\cite{HOHLER201919}, and we would expect it to fail as well here for the microgel packings in the high $\phi$, fully faceted, regime.  
More realistic and appropriate models take into account the nature of the deformation of the polymer network itself~\cite{ISI:000318945100010,ISI:000459588200002,ISI:000408781200082,ISI:000418665900001,doi:10.1073/pnas.2008076117,doi:10.1063/1.5026100}. 
Nikolova et. al.~\cite{doi:10.1073/pnas.2008076117} studied a bead-spring coarse-grained model with dissipative particle dynamics (DPD) to model both the elasticity of the gel network and the dynamics of solvent expulsion out of the network as the packing is confined.
They allowed their particles to swell and subjected them to isotropic osmotic confinement.  
They found that, above $\phi=1$, $K/K_p$ approached a constant value of about $0.8$, in agreement with the experiments of Liétor-Santos et. al. ~\cite{ISI:000297794700001}, but they did not study the shear modulus.

Here, we take a different approach.
Rather than using a bead-spring model to explicitly represent the gel network, we represent the gel network as a homogenized continuum using the Flory-Rehner constitutive law.
This approach is standard in the solid mechanic's community where various non-linear elastic properties of macroscopic swollen gels are of interest~\cite{ISI:000300400000020,ISI:000281620400014,ISI:000296170800001,ISI:000268500300015}, however, perhaps surprisingly, it has not been applied to particle packings.
The model can easily incorporate spatial variations in local cross-linking densities within a particle which may arise in various particle synthesis procedures (e.g. a hard, moderately swollen, core with high cross-link density enclosed by a soft, highly swollen, corona with low cross-link density), but in this preliminary work, we assume a homogeneous cross-linking density across the particle, and we assume the same cross-linking density for large and small particles.
The main disadvantage of our approach is that we cannot realistically study the fluid dynamics of solvent uptake/expulsion from the network and are restricted to compression and shear rates which are slow enough that we can assume the hydrodynamic forces generated by solvent flows are negligible.   
However, as we are primarily interested in the quasi-static regime in this work where the packing is sheared slowly enough to allow the fluid to be fully expelled/absorbed before further shearing, we are not adversely affected by this limitation.
Another potential disadvantage is that we are not able to capture inter-digitation of polymer segments across inter-particle facets~\cite{Mohanty:2017aa,doi:10.1073/pnas.2008076117,doi:10.1063/1.5026100,Scotti:2022aa}.
However, the effects of inter-digitation are largely uncontrolled in bead-spring models; this will give rise to tangential forces at the facets in a way that is not particularly well controlled or calibrated.
In our opinion, it is best to study and characterize a model which at first excludes these effects explicitly and only later introduces them in a controlled way.
Furthermore, in~\cite{doi:10.1073/pnas.2008076117}, the authors tune the properties of their bead-spring network to make sure that it behaves mechanically as a Flory solid, whereas here, we simply start with a Flory solid by construction.

Flory's constitutive law is governed by a total free energy density, $W_{tot}$, which gives the free energy per unit (unswollen reference) volume where $W_{tot}=W_e+W_m$ receives independent contributions from non-linear elastic deformation of the polymer network, $W_e$, and from the free energy of mixing of the polymer and solvent, $W_m$. 
We have, for $W_e$ and $W_m$~\cite{doi:10.1063/1.1723791,doi:10.1063/1.1723792,colbyBook,doiBook,ISI:000300400000020}, 
\begin{equation}
	W_e = \frac{Nk_bT}{2}[F_{\alpha \beta} F_{\alpha \beta} - d - 2 \log(J)]
\end{equation}
\begin{equation}
	W_m=\frac{k_bT}{\Omega}(J-1)\left[\log\left(\frac{J-1}{J}\right)+\frac{\chi}{J}\right]
\end{equation}
where $N$ is the number of cross-links per unit volume \emph{in the dry reference state}, $k_b$ is Boltzmann's constant, $T$ is the temperature, $F_{\alpha\beta}=\delta_{\alpha\beta}-\partial_\alpha u_\beta$ is the deformation gradient where $u$ is the displacement of the polymer network away from its unswollen reference configuration, $d$ is the dimension of the system, $J$ is the determinant of $F_{\alpha\beta}$ (so that $J$ gives the volumetric expansion at a given point relative to the dry reference and $J^{1/d}=\lambda$ gives the linear stretch ratio), $\Omega$ is the volume of a solvent particle (which is taken to be constant) and $\chi$ gives the energetic contribution to the free energy of mixing.  
In this preliminary work, we consider entropic swelling only and set $\chi=0$, so
\emph{there is one and only one dimensionless parameter in the model}: $N\Omega$, the number of cross-links in the dry reference configuration per unit solvent particle volume. 
For large $N\Omega$ the particles will be stiffer and undergo less free swelling, while for small $N\Omega$, the particles will undergo more free swelling and be softer.
If we consider the isotropic swelling case with a linear stretch factor, $\lambda$, we have, $F_{\alpha\beta}=\lambda \delta_{\alpha\beta}$ and $J=\lambda^d$ so $F_{\alpha\beta}F_{\alpha\beta}=d\lambda^2=d J^{2/d}$.
The osmotic pressure, $\Pi$, is simply the derivative of the free energy with respect to the logarithm of the volume, $\Pi=J\frac{\partial W}{\partial J}$, so
\begin{equation}
\frac{\Pi}{Nk_bT}=(J^{2/d}-1)+\frac{1}{N\Omega}(1+J\log[(J-1)/J])
\label{eq:piOfJ}
\end{equation}
The degree of free swelling is determined by setting $\Pi=0$ and solving for the equilibrium, $J_0$.
$J_0$ will depend on $N\Omega$, with larger $N\Omega$ (more cross-linking) giving less unconstrained swelling~\cite{doi:10.1063/1.1723791,doi:10.1063/1.1723792,colbyBook,doiBook,ISI:000300400000020}.
We study three different $N\Omega$ values: $1/20, 1/133, $ and $1/754$, and solving for $\Pi=0$ gives: $J_0=4.00762, 8.99301, 20.2519$ respectively.
That is: our least densely cross-linked (softest) particles freely swell to roughly $20$ times their dry area, while our most densely cross-linked (hardest) particles freely swell to roughly $4$ times their dry area.


We simulate the system using standard finite element method (FEM) techniques using constant strain triangle (CST) elements to mesh the particles~\cite{Hutton2003FundamentalsOF} and use a simple gradient descent to re-equilibrate the system.
We use a 50:50 mixture of two species of circular particles with the ratio between the radii equal to $1.4$~\cite{ISI:000184582400020}.
The disks are initialized on a square grid, allowed to swell freely, and then slowly compressed.

To enforce impenetrability constraints, we simply introduce a strong power-law repulsion between surface nodes of opposing particle FEM meshes: $V_{ij}=V_0(r_{ij}/R)^{-a}$ where $r_{ij}$ is the distance between the surface nodes on opposing particle meshes, $R$ is the length-scale of the repulsion, $V_0$ is the energy scale of the repulsion, taken here to simply be $NkT$, and $a=12$ is an exponent which is chosen to be large enough that it prevents inter-penetration and does not otherwise affect the results.
To avoid artificial interlocking of surface nodes and the resulting tangential forces, we choose a lengthscale for the power-law repulsion which is roughly $6$ times the surface node spacing in the dry reference mesh, and we find that this sufficiently suppresses tangential traction forces at the facets. 
Because of the stiffness of the surface repulsion, the surface interactions do not contribute significantly to the overall energy or stress, so we neglect them when reporting stresses and moduli.
However, the numerical scheme results in a skin at the facets between the surface nodes which has a constant size which does not change as the packing is subjected to contraction of the space.
This results in a larger and larger fraction of the space being taken up by the spurious skin as the system is compressed.
To compensate for this artifact, we define the Cauchy stress, whose trace is the pressure, as the derivative with respect to an infinitesimal strain increment of the total energy of the deformed elastic network divided by the area occupied by the gel \emph{excluding the area of the skin region}.  
We have checked that the contribution of the power-law repulsion to the total energy is negligible as long as the exponent is chosen to be sufficiently large.
We similarly define $\phi=J_0/\langle J\rangle$ as the ratio of the total freely-swollen particle area to the current total particle area occupied \emph{excluding the skin}.
This definition of $\phi$ -- which is, by construction, greater than or equal to $1$ for compressed particles -- is somewhat un-natural near the jamming onset where one would want to divide the freely swollen particle volume by the area of the simulation cell rather than just the area occupied by the polymer network.
To measure $\phi=J_0/\langle J\rangle $ in an experiment, one would need to have an independent measurement of the pore volume in the sample to infer the current total particle volume, but, nevertheless, it is the natural quantity to use in the model to make a direct connection between the mechanical response of the packing and the mechanical response of the equivalent monolithic continuum. 

\section{Results}
\begin{figure}[tb]
\includegraphics[width=0.99\columnwidth]{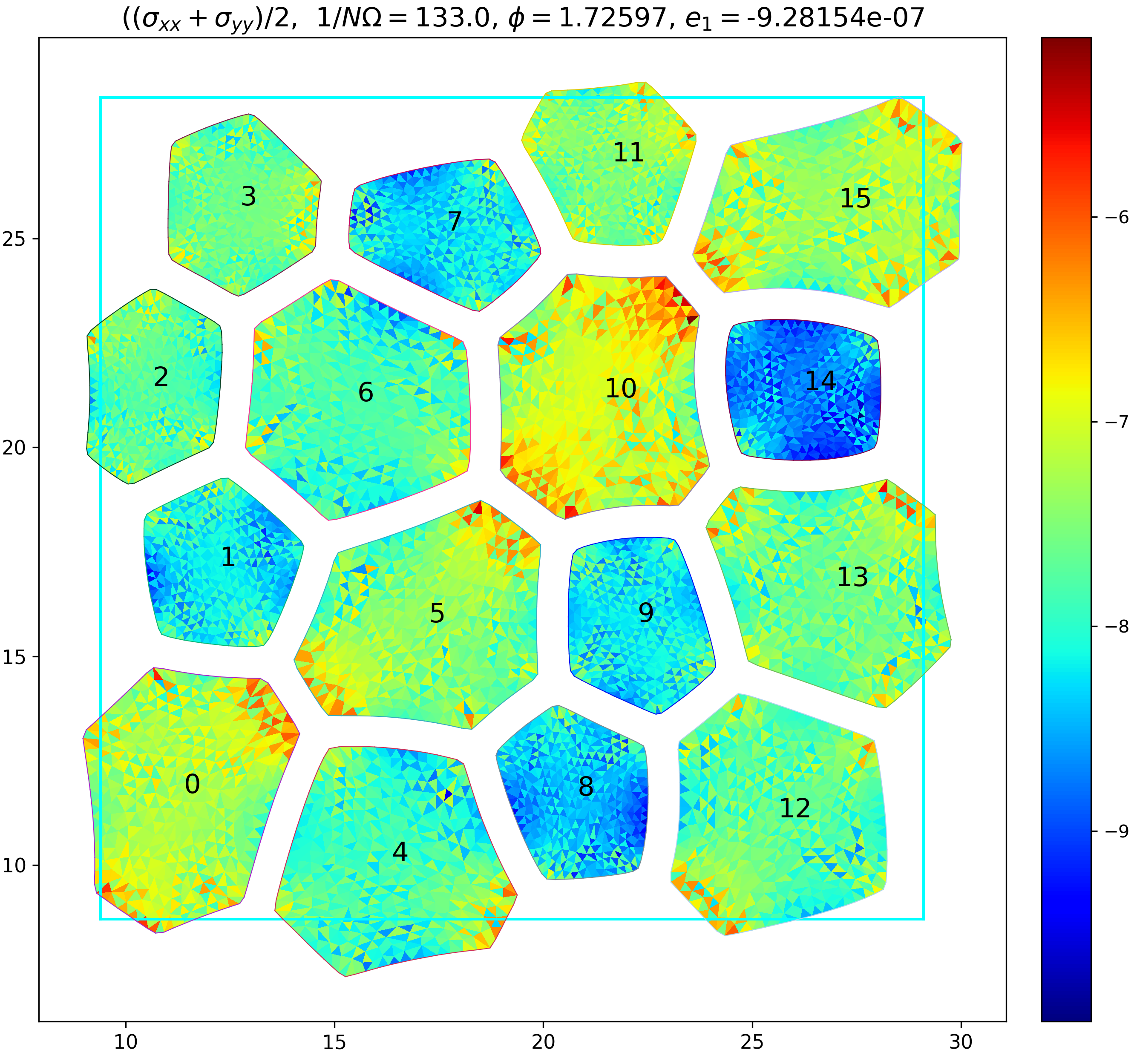}
\caption{
Local pressure, $\Pi$, in units of $NkT$ for a typical system of $16$ particles with $N\Omega=1/133$ at a nominal $\phi=1.73$.
Particles are labeled with a unique identification number for discussion in the text. 
}
\label{sfig:pressureImage}
\end{figure}
In figure~\ref{sfig:pressureImage}, we show an image of the local pressure in units of $NkT$ for a typical system of $16$ particles with $N\Omega=1/133$ at a nominal $\phi=1.73$ which is in the fully faceted regime with no remaining voids of pure solvent.
Each triangular element in the FEM mesh is colored according to its pressure (with the standard solid mechanics convention that compressive stresses are negative). 
We observe some high-frequency oscillations from element to element, but we have checked that these are suppressed using higher-order finite element schemes and do not affect any of the main results of this work.
Some general trends can be seen here which are observed in other members of the ensemble and at various other $N\Omega$ and various other $\phi$.
i) Small particles tend to have higher pressure than large particles.
ii) Pressure tends to be larger near the centers of facets than near the facet junctions.
iii) Facets between large and small particles tend to be convex on the small-particle side and concave on the large-particle side: that is, the small particles tend to protrude into the large ones.
iv) Particles with fewer facets tend to be at larger pressure and vice versa.  For instance, particle $14$, a small particle which has four neighbors, is at a larger pressure than the other small particles which have $5$ or $6$ neighbors, while particle $10$, a large particle which has eight neighbors, is at a smaller pressure than the other large particles which have $7$ or $6$ neighbors.

\begin{figure}[tb]
\includegraphics[width=0.95\columnwidth]{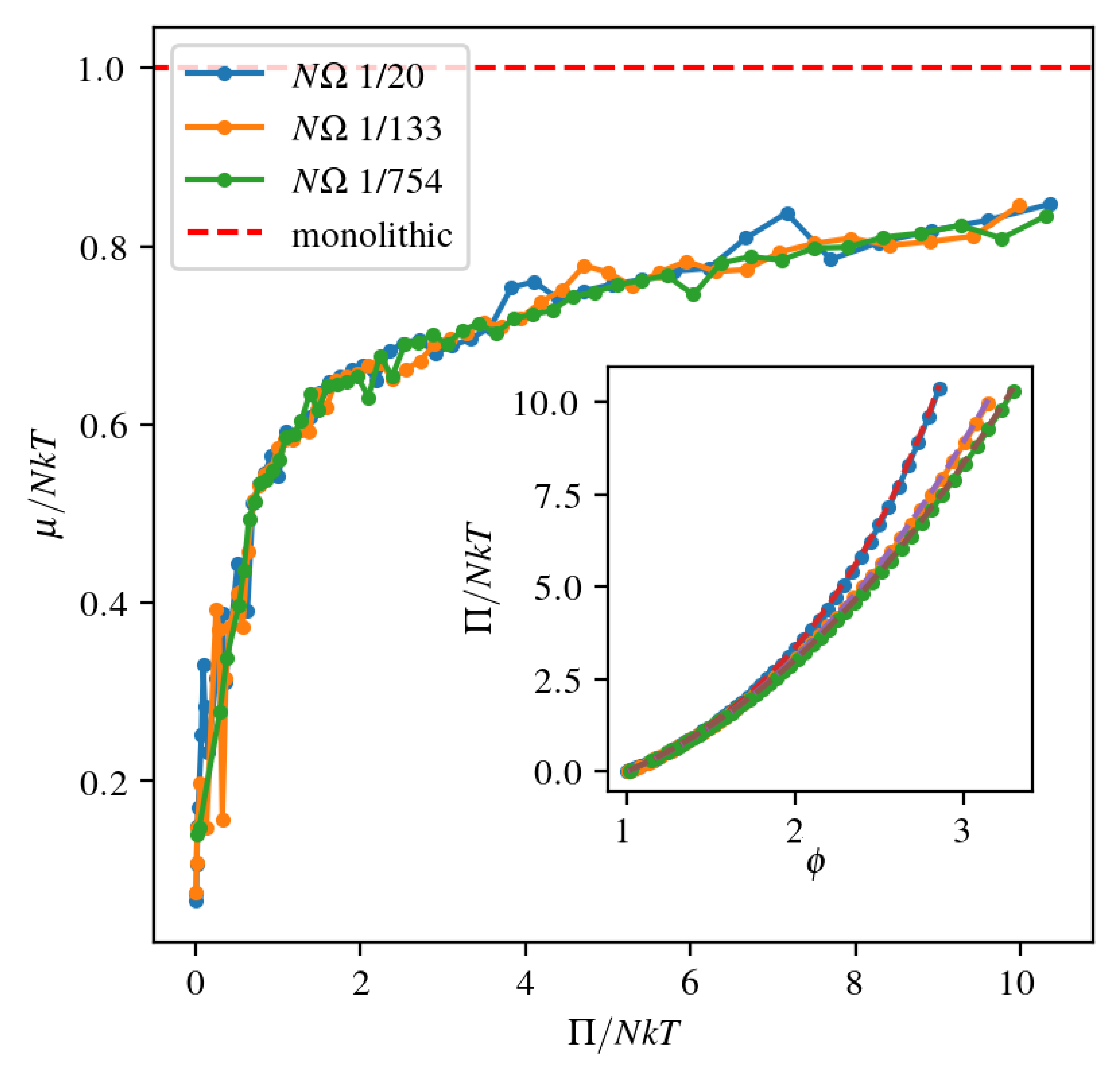}
\caption{
Main: Shear modulus vs. osmotic pressure, $\Pi$, for various cross-link densities. The red line is the shear modulus of a two-dimensional monolithic Flory material.  
Inset: $\Pi/NkT$ vs $\phi$.  Dots are data from the FEM simulation.  Solid lines are for a monolithic Flory solid. 
}
\label{sfig:mu_vs_phi}
\end{figure}

In the inset of figure~\ref{sfig:mu_vs_phi}, we show the pressure, $\Pi$, vs $\phi$ (recall the definition of $\phi=J_0/J$ excludes regions of pure solvent).
The results are for an ensemble of $36$-particle systems.
The symbols represent the pressure computed in the actual packing, and the curves simply represent the pressure for an isotropic monolith of Flory material as computed from equation~\ref{eq:piOfJ} at the given $J=J_0/\phi$.
The agreement is striking!
The pressure field inside the particles is inhomogeneous: it is large near the facets and smaller near the vertices in the facet network with overall fluctuations of order tens of percent.
Yet despite this inhomogeneity, the \emph{spatial average} of the pressure is very close to that of the isotropically confined Flory monolith at the same density as the \emph{average density} of the packing.
There is no fundamental reason why this relation must be an exact identity, and we do see deviations from the monolithic result of the order of a few percent.
It is a manifestation of the fact that the distribution of local $J$ values has a width that is small relative to its average and that the $\Pi (J)$ function from equation~\ref{eq:piOfJ} varies relatively slowly on that scale so that $\langle\Pi(J)\rangle\approx \Pi(\langle J\rangle)$.

The pressure diverges locally as $J\rightarrow 1$ from above, and this sets an upper bound on $\phi$ at $\phi_{\text{max}}=J_0$ as the solvent is completely expelled from the system.
Accordingly, we see $\Pi$ begin to diverge first for the system with the highest $N\Omega$ and last for the system with the lowest $N\Omega$.
Our numerical scheme becomes unstable at a pressure of about $\Pi=10Nk_bT$ for all systems, so we are able to reach slightly higher $\phi$ for the systems with lower $N\Omega$.

To measure the shear modulus, we make a small axial deformation to all FEM nodes and the periodic boundaries with extension along the horizontal, $x$, and contraction along the vertical, $y$: $x\rightarrow e^{\epsilon} x\approx (1+\epsilon)x, y\rightarrow e^{-\epsilon}y\approx (1-\epsilon)y$, re-equilibrate, and then measure the tangent modulus as $\mu=\Delta\sigma/\epsilon$ where $\Delta\sigma$ is the change in the shear component of the Cauchy stress.
Since we hold the cell in a square shape during the initial swelling of the circular particles, there will be a random residual shear stress which is distributed normally about zero in our ensemble.
It is very small compared to $\mu$ and does not affect the results.

In the main plot of figure~\ref{sfig:mu_vs_phi}, we show the shear modulus, $\mu$, vs the osmotic pressure, $\Pi$.
For a monolithic Flory material in 2D, $\mu=Nk_bT$ regardless of $N\Omega$ and completely independent of $\Pi$ (and/or $\phi$).
For our particle packing, the curves all start at zero near the jamming point at $\Pi=0$ and increase monotonically remaining below the monolithic Flory value.
Strikingly, the curves collapse, indicating that the shear modulus of the packing is a function of the pressure alone and is independent of the cross-linking density (after scaling by $NkT$).
This observation for the shear modulus is in a similar spirit to the observation of  Liétor-Santos et. al. ~\cite{ISI:000297794700001} that the pressure and shear modulus scale like the single-particle compression modulus, but must differ in detail as we argue below.

\begin{figure}[tb]
\includegraphics[width=0.85\columnwidth]{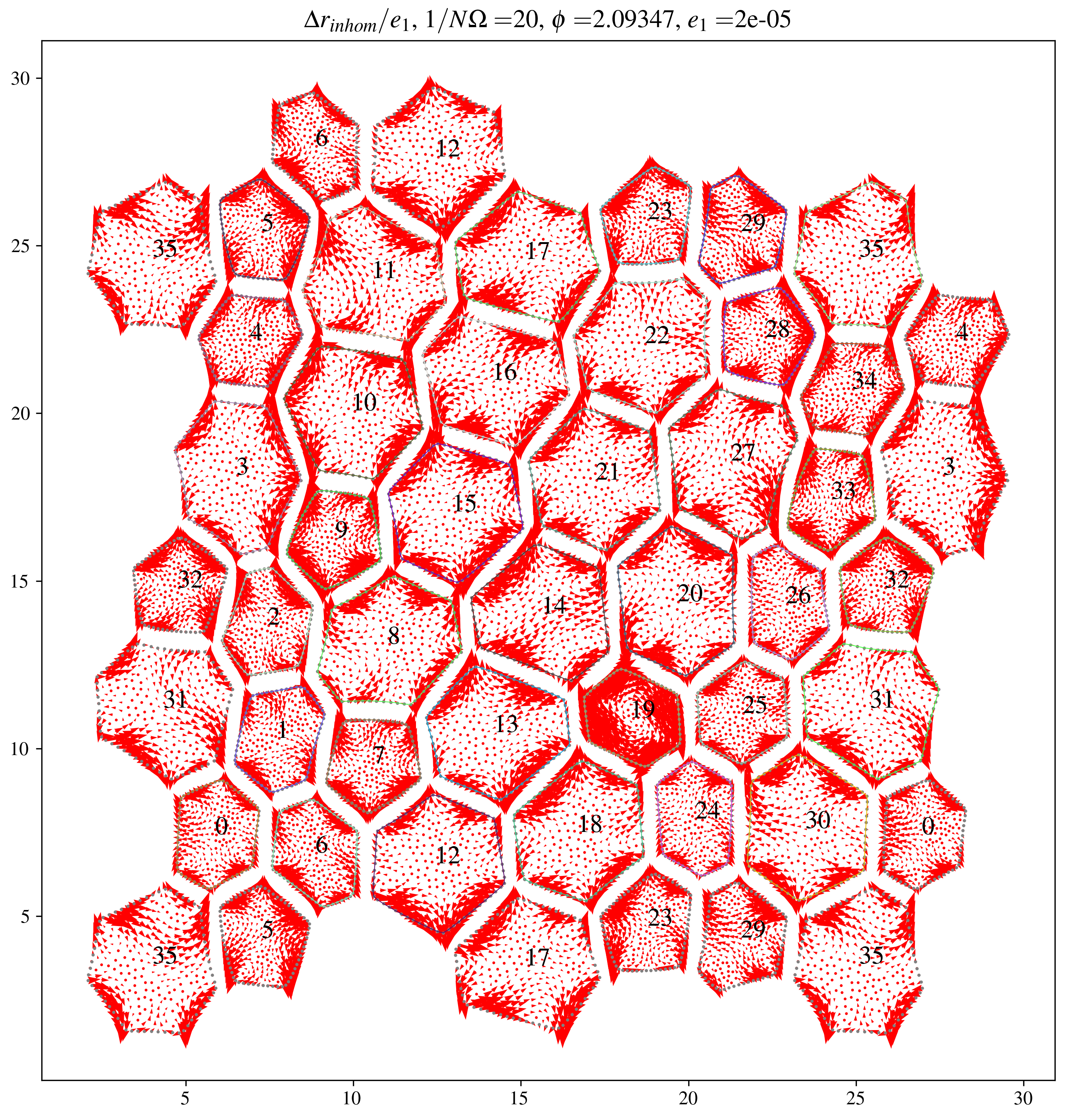}

\includegraphics[width=0.85\columnwidth]{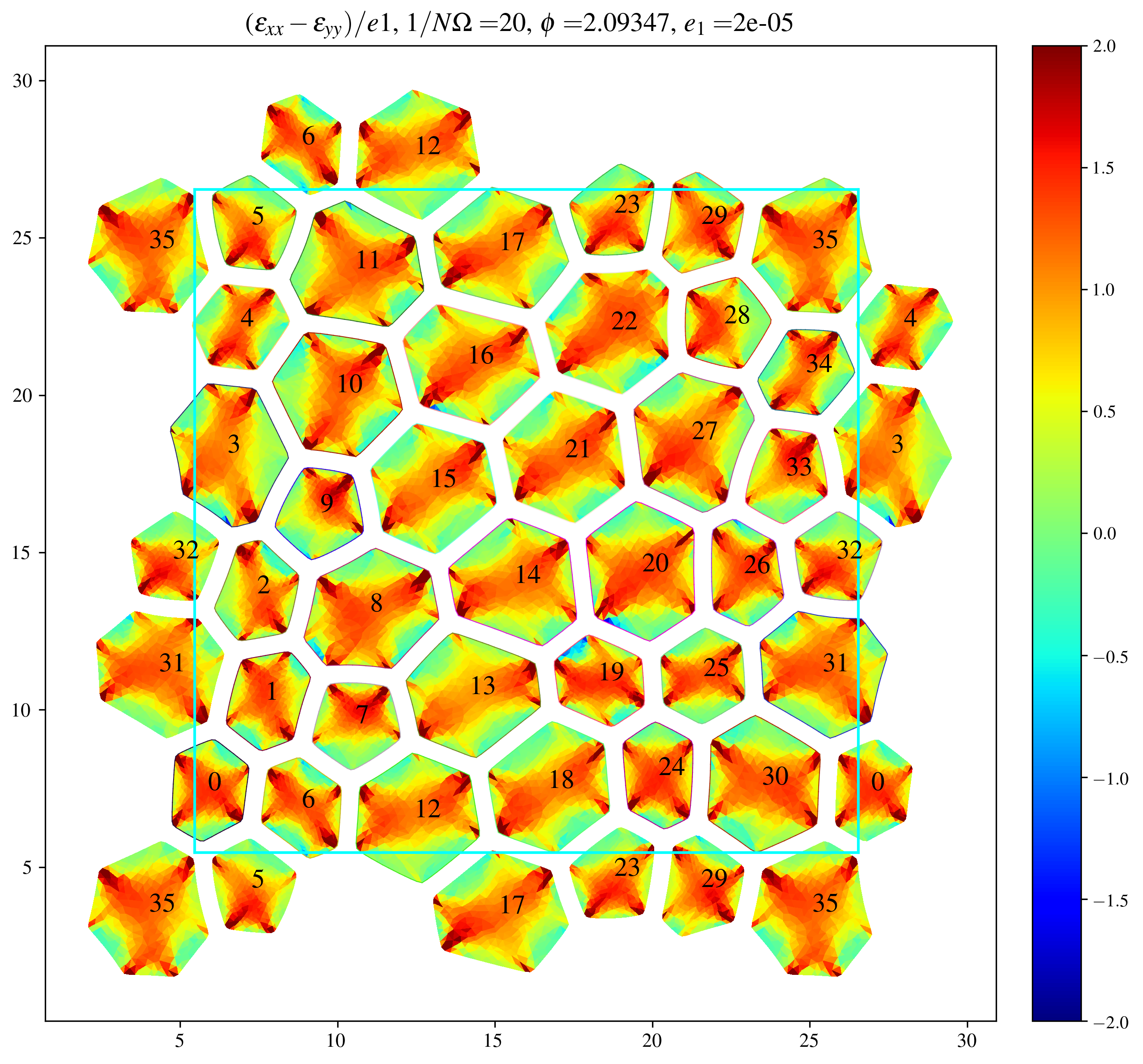}

\caption{(a) Non-affine displacement field normalized by the imposed shear strain.
The imposed shear strain is axial with vertical contraction (horizontal extension) and area-preserving at $\phi$ = 2.1 for a system with cross-linking density $N\Omega$ = 1/20. 
Arrows are drawn to scale to precisely give the non-affine displacement per unit applied strain.
(b) Local shear strain, $(\partial_{x}u_x-\partial_y u_y)/2$, scaled by the imposed shear strain, $e1$.
}
\label{fig:slip_figures}
\end{figure}

In figure~\ref{fig:slip_figures} (a), we show a typical $N\Omega=1/20$ system with $36$ particles at $\phi=2.09$ subjected to a small strain step as described above.
We show only the non-affine component of the nodal displacements. 
In figure~\ref{fig:slip_figures} (b), we show the combination of the components of the gradient of the displacement corresponding to the applied strain: $(\partial_x u_x - \partial_y u_y)/2$ normalized by the applied strain so that a value of $1$ indicates that the material is locally shearing precisely according to the globally imposed shear strain.
The strain is quite inhomogeneous across the particles.
In the centers of all the particles, the shear strain is approximately equal to the globally imposed shear strain, while the deformation near facets and vertices depends on the orientation.
Facets that are roughly vertical or horizontal have essentially no relative displacement, while facets that are roughly diagonal across the cell have a significant amount of relative tangential motion indicating sliding along those facets.   

\begin{figure}[tb]
\includegraphics[width=0.95\columnwidth]{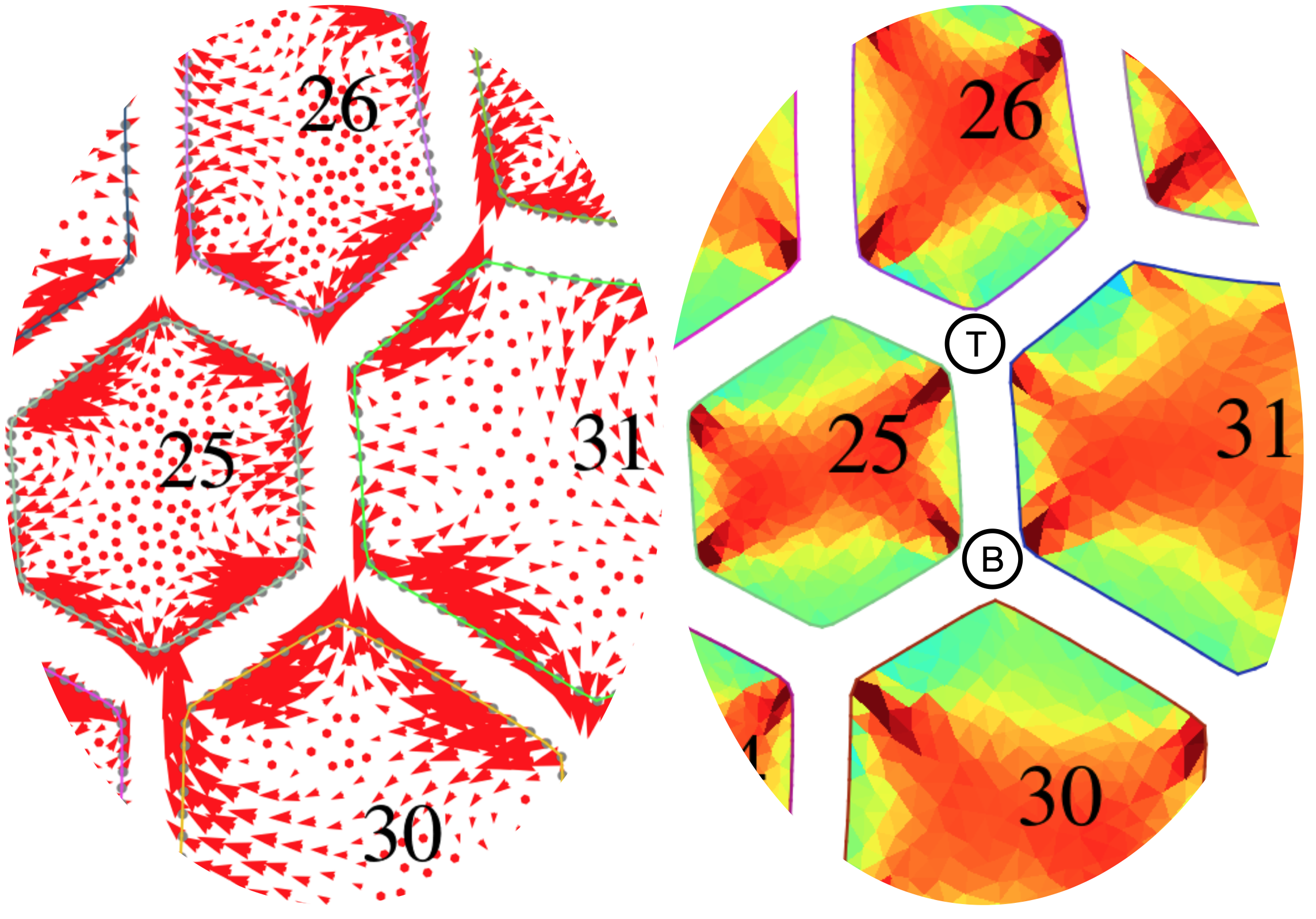}
\caption{Closeup on the facet between particles $25$ and $31$ from figure~\ref{fig:slip_figures}.
The triple junction with particle $26$ on the top and $30$ on the bottom are labeled $T$ and $B$ respectively for ease of discussion in the text.
}
\label{fig:slip_figures_closeup}
\end{figure}
In figure~\ref{fig:slip_figures_closeup}, to illustrate this point, we zoom in on the facet between particles $25$ and $31$, since this facet is nearly vertical.
For ease of discussion, we label the triple junction with particle $26$ on top and with particle $30$ on the bottom as $T$ and $B$ respectively. 
Since the facet is a contact between a large and small particle, it has a slight curvature toward the small particle, but this should not significantly affect the response.
The facet is nearly vertical, and there is essentially no non-affine displacement on either side.
However, the displacement on particle $26$ at the $T$ vertex is downward, and the displacement on particle $30$ at the $B$ vertex is upward, so the facet is \emph{shortening}.
As particles $26$ and $30$ are advancing into the facet, the strain near their vertices at the triple junctions is nearly zero: green in the color scheme of the image.  
On the other hand, the material near the four vertices on particles $25$ and $31$ at the $T$ and $B$ triple junctions is very strongly sheared -- with local shear strains greater than the applied shear strain -- to accommodate the shortening of the facet. 
The two other facets connected to $T$ and the two other facets connected to $B$ are all oriented approximately $30$ degrees from horizontal and have a significant amount of slipping.
Since the tangential traction forces at the slipping interface are zero, and the shear stress must be zero along the facet, the slip results in a screened region where the strains in the particles are quite small and nearly zero at the facet.

The same scenario plays out in reverse for facets which are oriented horizontally.
The result is that all particle vertices in the packing whose opposing facet is nearly horizontal or vertical should have a very low strain in their neighborhood, and a quick scan of the packing shows this to be true.
For example, the facet between particles $34$ and $35$ is nearly horizontal, so it is lengthening, and the corresponding vertices on particles $28$ and $4$ have a very low shear strain on them: the fact that particle $28$ has $5$ neighbors and is compressed into a pentagon is incidental and has little impact on this result.





\begin{figure}[!tb]
\includegraphics[width=0.45\columnwidth]{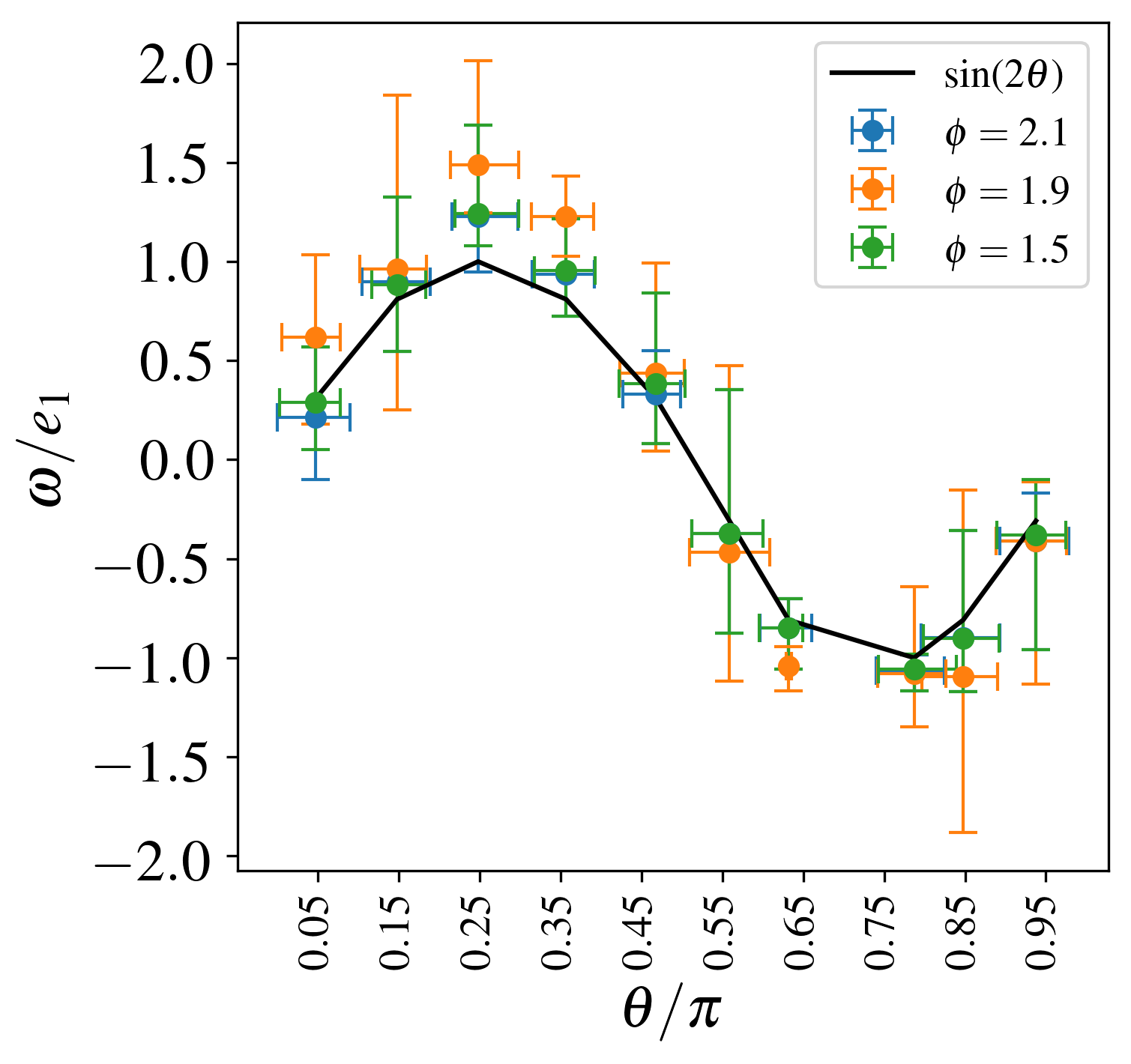}
\includegraphics[width=0.45\columnwidth]{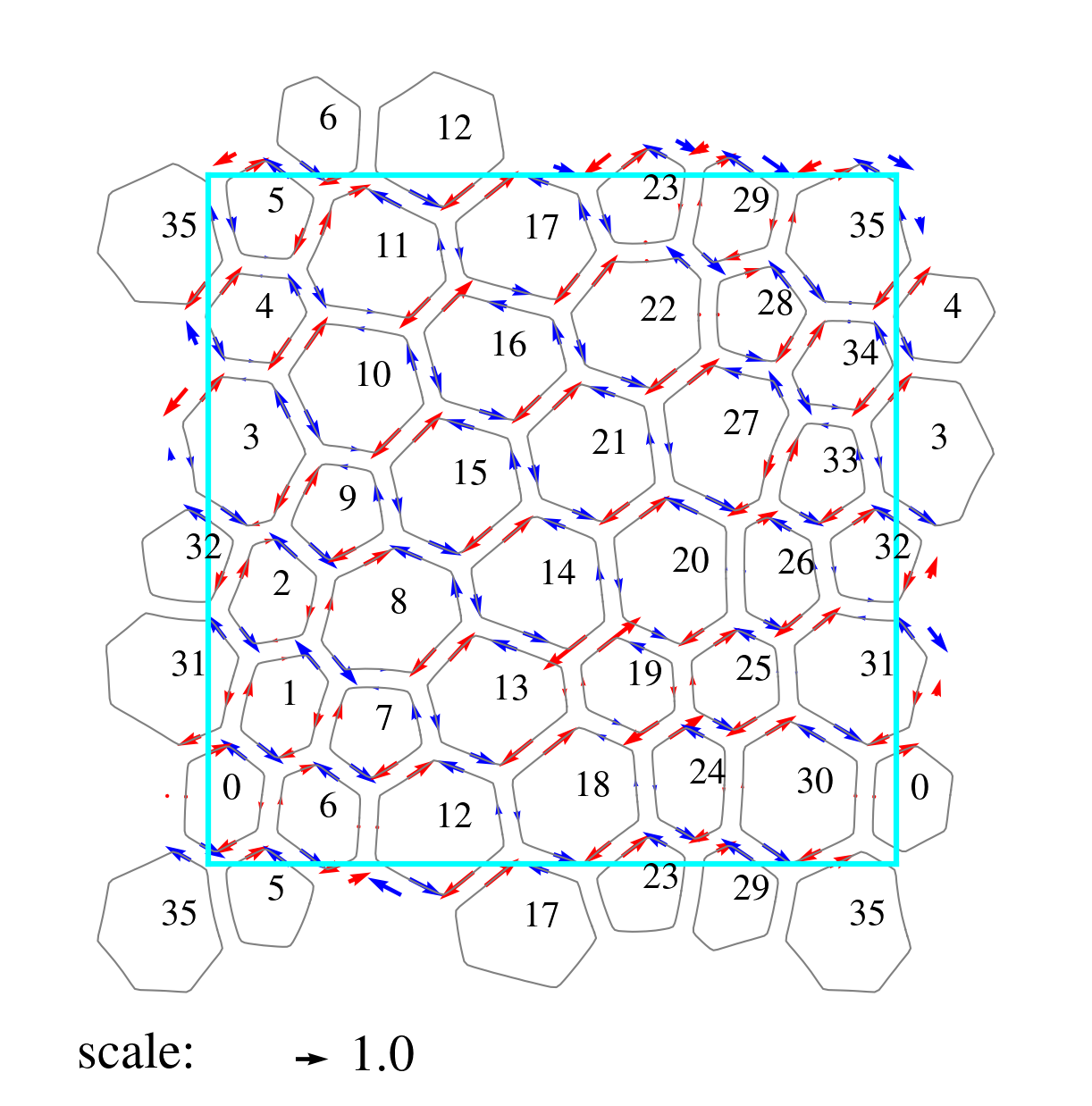}
\caption{
a) Facet slip , $\omega$ (defined in text) scaled by applied strain $\epsilon$ vs. facet angle $\theta$ for three different $N\Omega$ systems at $\phi=1.5, 1.9$ and $2.1$. 
b) Facet slip in real space with arrow length indicating the amount of slip with red and blue indicating counter-clockwise and clockwise vorticity respectively.}
\label{sfig:facet_slip}
\end{figure}

\begin{figure}[!tb]
\includegraphics[width=0.45\columnwidth]{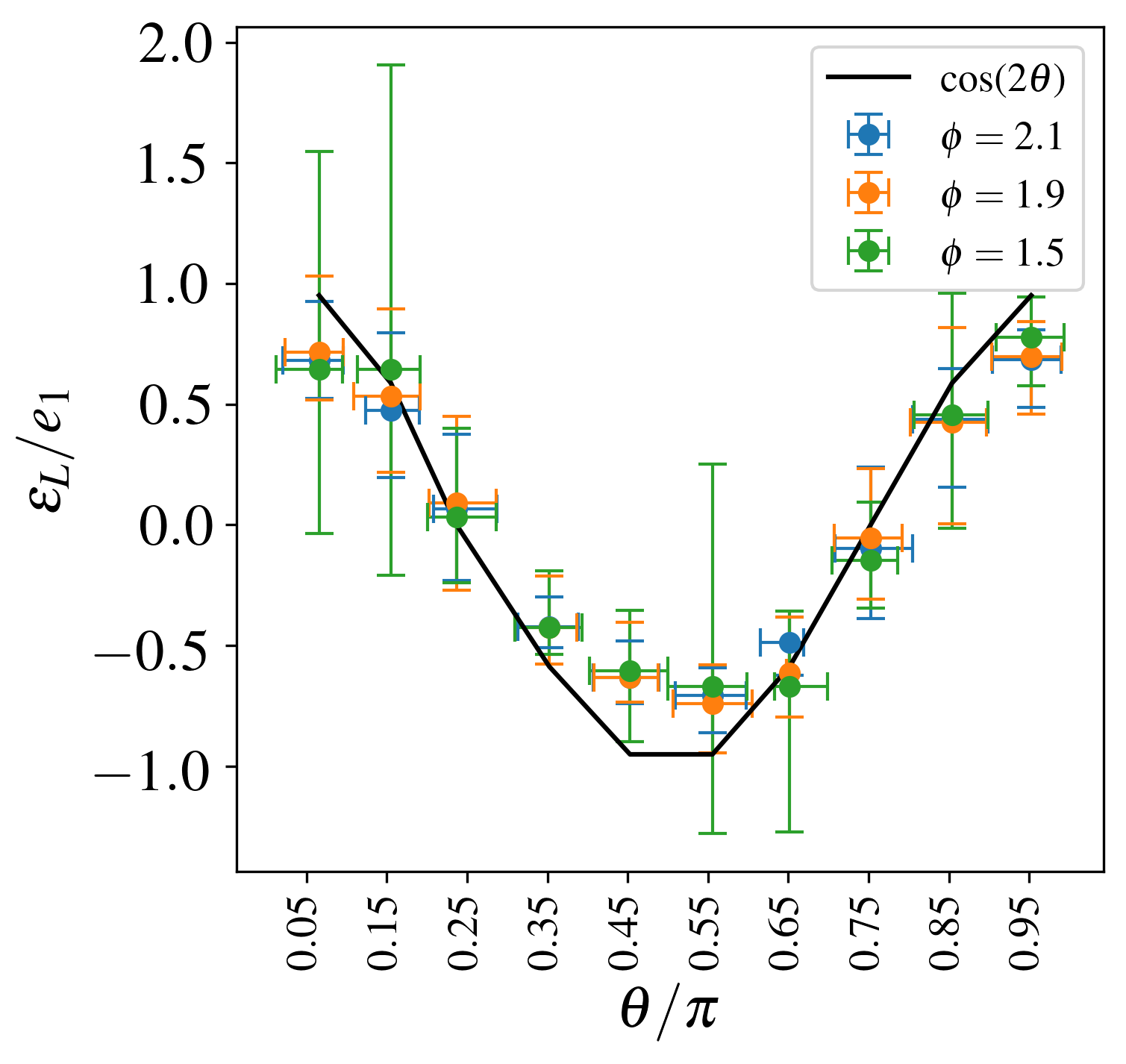}
\includegraphics[width=0.45\columnwidth]{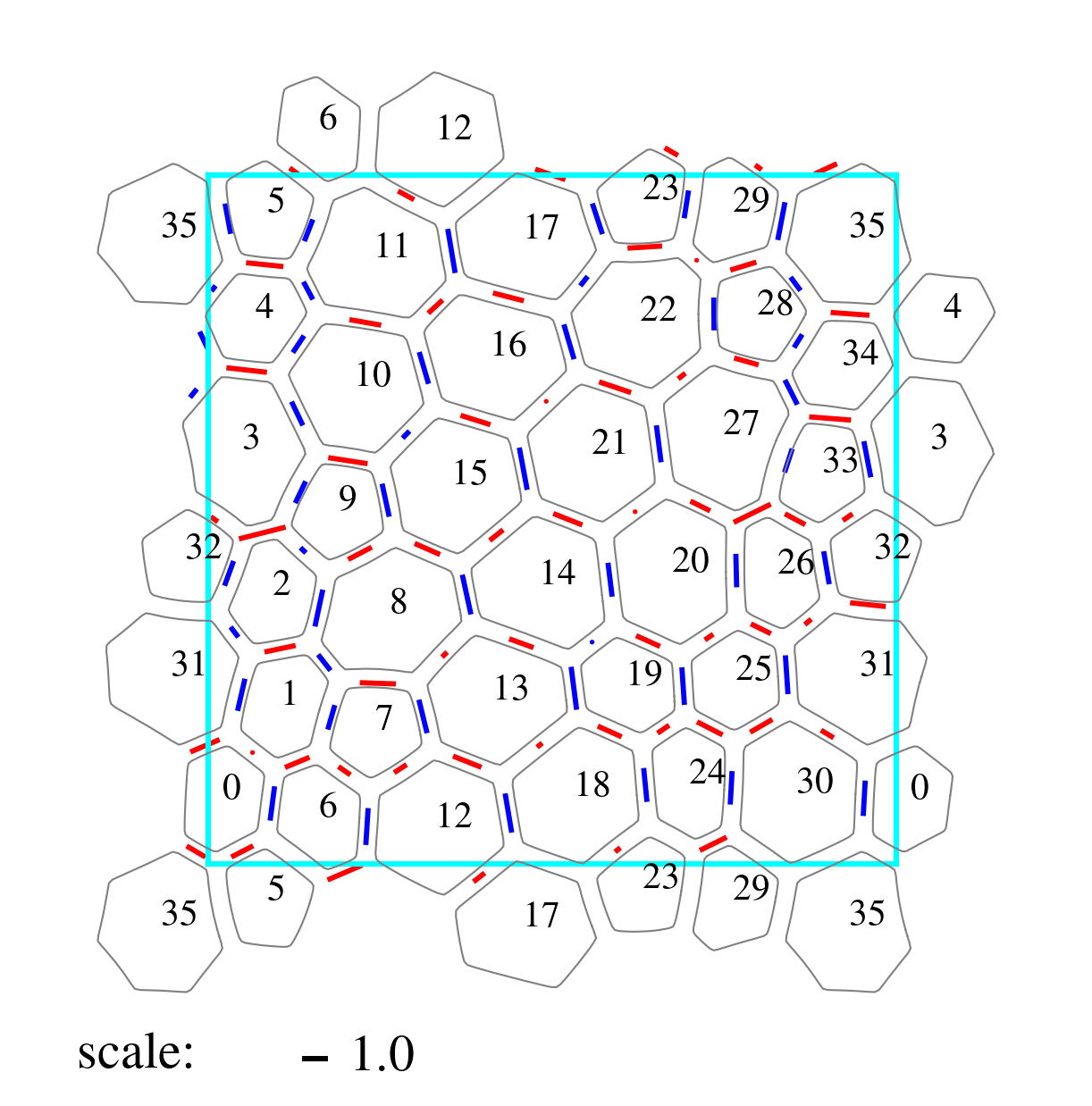}
\caption{a) Facet strain, $\epsilon_L$ (defined in text) scaled by applied strain $\epsilon$ vs. facet angle $\theta$ for three different $N\Omega$ systems at $\phi=1.5, 1.9$ and $2.1$. 
b) Facet strain in real space with bar length indicating the amount of relative lengthening (red) or shortening (blue).}
\label{sfig:facet_strain}
\end{figure}
In figures~\ref{sfig:facet_slip} and ~\ref{sfig:facet_strain}, we show the slip and strain on the facets as a function of the facet angle.
We approximate the facet network as polygonal and identify the vertices of the polygonal network with the surface nodes of the mesh located at the triple junctions. 
Then, to define the facet slip, $\omega$, we simply take the difference in the average tangential component of displacement on either side of the facet.
Similarly, to define the facet strain, $\epsilon_L$, we take the difference in the component of the end-to-end displacement along the end-to-end separation and divide by the current facet length. 
What we see agrees remarkably well with the anecdotal description of the displacement fields in figure~\ref{fig:slip_figures} where facets along the box diagonals slip significantly but neither shorten nor lengthen while facets along the axes do not slip but shorten (the vertical ones) or lengthen (the horizontal ones).
In fact, we can make some simple assumptions about the deformation kinematics.
If we assume the slip on a facet is equal to the locally resolved transverse component of the applied displacement gradient: $\omega=\hat{n}_\alpha \hat{t}_\beta \partial_\alpha u_\beta$ where $\partial_\alpha u_\beta$ is the imposed deformation and $\hat{n}$ and $\hat{t}$ are the unit normal and tangent to the facet, then we would get: $\omega=e_1 \sin(2\theta)$ where $e_1$ is the amplitude of the imposed strain and $\theta$ is the angle of the tangent. 
For the strain, if one imagines that the triple junctions in the facet network deform affinely with the imposed shear, one would expect that $\epsilon_L=\hat{t}_{\alpha}\hat{t}_{\beta}\partial_\alpha u_\beta = e_1\cos(2\theta)$.
We see that the data follow that trend remarkably well regardless of $\phi$ with no discernible trend with $\phi$.
Our argument seems to slightly overestimate the slip and underestimate the strain, but all things considered, this prediction with no adjustable parameters seems to work out well: the deformation of the facet network is essentially affine and equal to the homogeneously imposed deformation.

\begin{figure}[!tb]
\includegraphics[width=0.75\columnwidth]{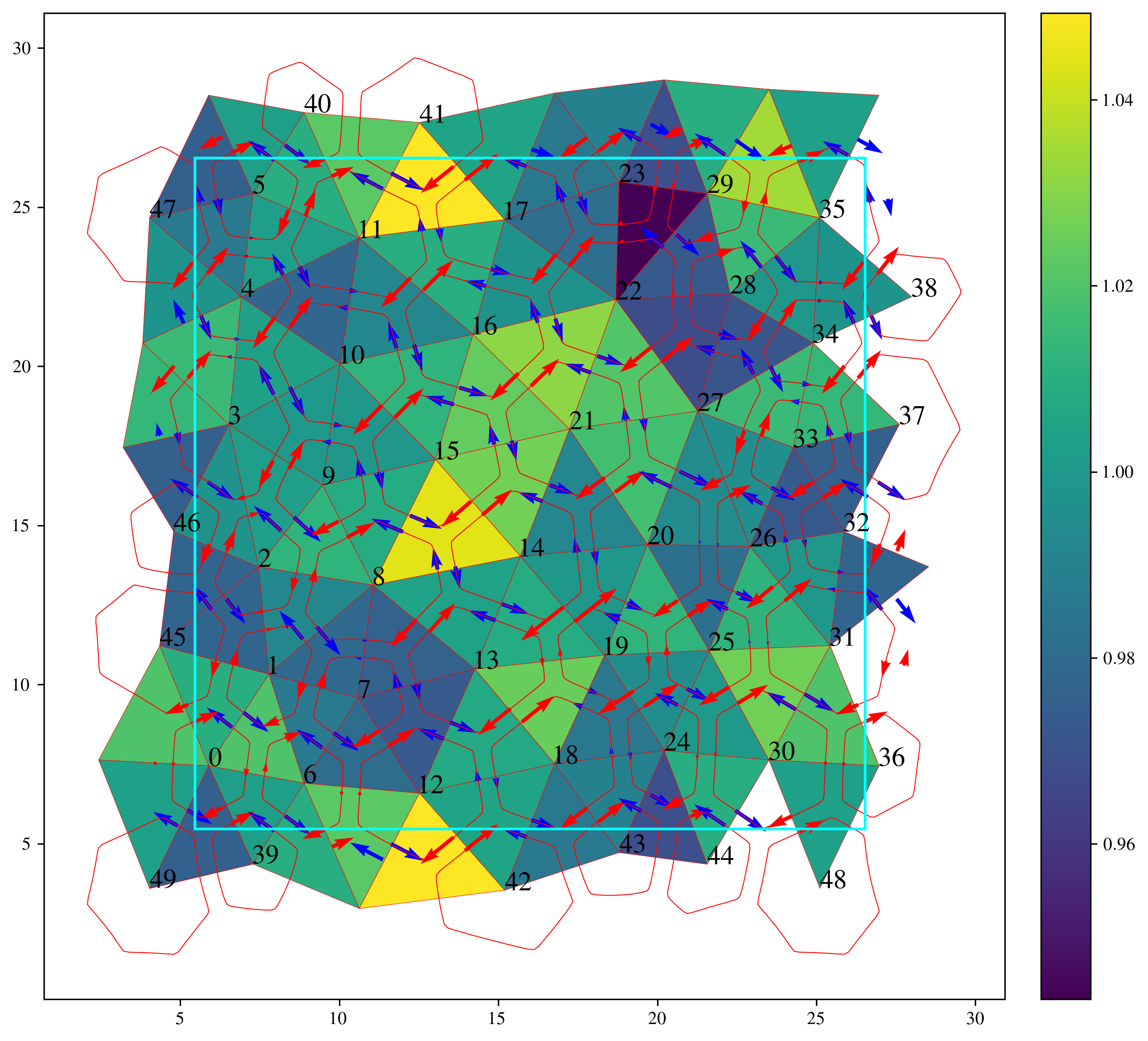}
\caption{Shear strain of the particle centroids scaled by the applied strain.  
The strain is computed for each Delaunay triangle in the contact network by linear interpolation. 
Slip vectors are reproduced from the figure for convenience~\ref{sfig:facet_slip}.
}
\label{sfig:Centroid_strain}
\end{figure}
In figure~\ref{sfig:Centroid_strain}, we show the strains associated with the motion of the particle centroids.
We define the strain as piecewise constant on the Delaunay triangulation~\cite{CompGeoBook} of the contact network using linear interpolation of the three centroid displacements.
What we see is that the resulting deformation of the particle centroids is quite homogeneous despite the large inhomogeneities \emph{within} any given particle.
It is hard to draw any statistical trends or make correlations between the Delaunay strain and any obvious geometrical or topological properties of the packing.

\section{Discussion and summary}


\begin{figure}[tb]
\includegraphics[width=0.45\columnwidth]{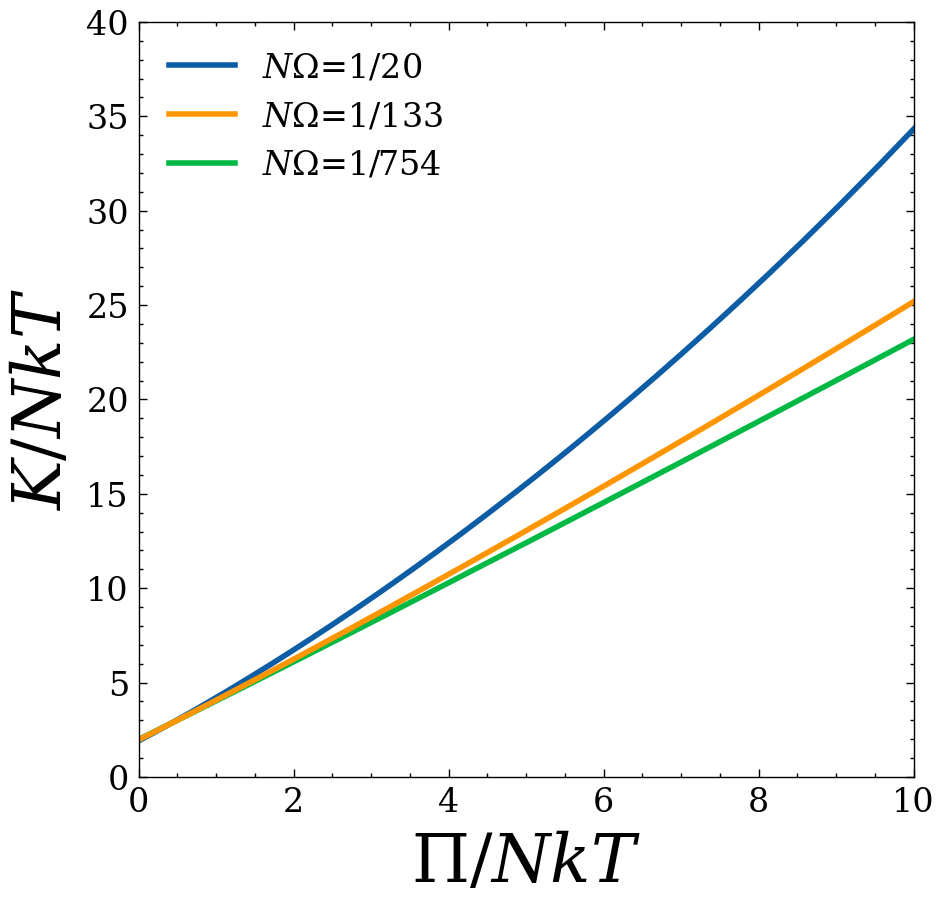}
\includegraphics[width=0.45\columnwidth]{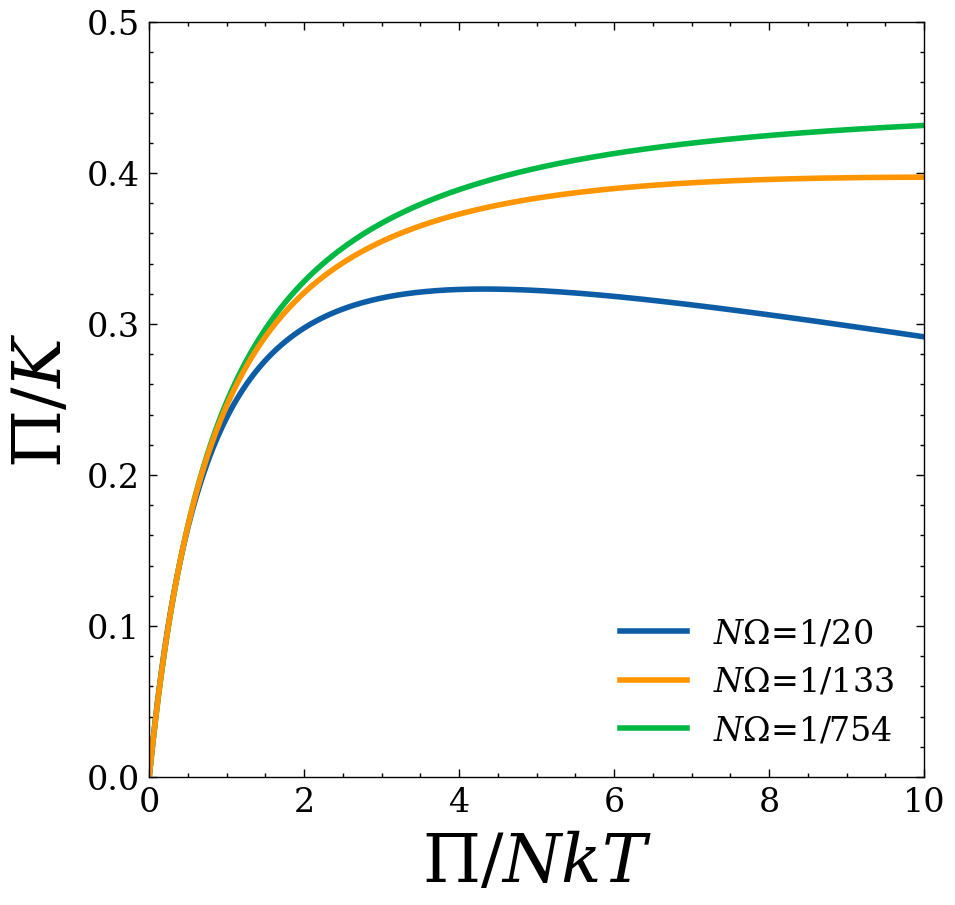}
\caption{
a) Compression modulus, $K/NkT$, vs. osmotic pressure, $\Pi/NkT$, for a monolithic Flory material in the range of pressures of interest in this work $0<\Pi/NkT<10$.
b) Their ratio, $\Pi/K$, vs $\Pi/NkT$.}
\label{sfig:kOverPi_vs_Pi}
\end{figure}
We have shown here that the microgel packings behave in one of the simplest ways one could have imagined: the pressure has a density dependence as if it were a monolithic Flory material, while the shear modulus, although it has a non-trivial $\phi$ dependence and starts from zero at the jamming point, is a universal function of the dimensionless pressure, $\Pi/NkT$, \emph{independent of cross-linking densities}. 
The $\mu/NkT$ vs $\Pi/NkT$ curve shows a transition from a more strongly pressure-dependent regime at low pressure near jamming onset to a much weaker pressure dependence in the fully-faceted regime and is consistent with the conjecture in Cloitre et. al.~\cite{ISI:000183582800003} that the transition to the weaker concentration dependence of $G$ occurred at the onset of full-faceting where solvent-pure void space completely disappeared.
This is also completely consistent with the arguments of Menut et. al.~\cite{ISI:000297561400020} who claimed that their systems showed monolithic Flory behavior at high density (although we point out that the  scaling behavior for a Flory monolith in $3D$ is $\mu\sim \phi^{1/3}$ rather than the $\mu\sim \phi^1$ suggested in that work~\cite{colbyBook,doiBook}

At face value, our results would seem to be inconsistent with Liétor-Santos et. al~\cite{ISI:000297794700001} where $\Pi$ became independent of $\phi$ when scaled by the particle compression modulus, $K_p$, rather than the Flory modulus, $NkT$.
However, in figure~\ref{sfig:kOverPi_vs_Pi}, we plot the compression modulus, $K$ vs. the pressure, $\Pi$, for our three $N\Omega$ values for a homogeneously compressed Flory solid where $K=J\frac{\partial \Pi}{\partial J}$.
The compression modulus for the zero pressure freely swollen state is around $K_{eq}\approx 2NkT$ and has only a weak dependence on cross-linking density $N\Omega$.
We see that $\Pi/K$ does not vary dramatically over the pressure range, and for the two more weakly cross-linked systems, $\Pi/K\approx 0.4$ above about $\Pi=4NkT$.
Since our $\Pi$ vs. $\phi$ curves for the packing are virtually identical to the monolith, this means that our results are roughly consistent with reference~\cite{ISI:000297794700001} where $\Pi/K$ was roughly constant above $\phi=1$.
Our results for the shear modulus, in distinction with the pressure and compression modulus, are not as easy to reconcile with reference~\cite{ISI:000297794700001}.
If we were to scale the shear modulus by $K$ rather than $NkT$, because of the $N\Omega$ dependence of $K$ illustrated in figure~\ref{sfig:kOverPi_vs_Pi}, we would break the good collapse obtained in figure~\ref{sfig:mu_vs_phi}.
Furthermore, the $\phi$-independent values of $G/K_p$ obtained in~\cite{ISI:000297794700001} are of the order of a few percent, whereas here, our values of $\mu/NkT$ are around $0.8$ for most of the pressure range of interest; our values for $\mu/K$ would vary much more strongly with pressure than our $\mu/NkT$ do.

What we've shown here is that, once the system is in the fully faceted regime, the kinematics of shear deformation becomes largely insensitive to the compression.
The vertices in the facet network, on average, follow the imposed homogeneous deformation.
The resulting slip at the facets results in a reduction of the modulus of the packing away from the monolithic Flory value by some tens of percent but remains of the order of the monolithic value.
In the future, it would be important to use the kinematic information about the affine motion of the facet-vertex network to try to construct quantitative estimates for the modulus reduction away from the Flory value.
Since each particle is assumed to be homogeneous, the result of imposing a homogeneous deformation on the particle's boundary would be a homogeneous deformation of the particle interior, and the result would be a recovery of the full Flory shear modulus as if the particles were welded together at the facets.
So it seems, going beyond considering the average kinematic behavior of the facet network would be necessary to obtain any corrections to the Flory shear modulus.

In this work, we made many simplifying assumptions.
We assumed that: i) the particles were homogeneous disks, ii) the free energy of mixing was completely entropic with $\chi=0$, iii) the facets were free of friction and neglected any possible effects of inter-digitation of the polymer networks across the facets, iv) the deformation was slow enough that kinetics and viscous flow effects were negligible and that the solvent was able to freely flow into and out of the polymer network with no resistance v) the particles were large enough that Brownian effects were negligible, vi) the system was 2D.
It will be useful to go beyond these assumptions in future models.
It would also be important to know how the many-body interactions studied here at large volume fraction affect the plastic yielding behavior at large strains and the glass transition for the Brownian case, and, more generally, the overall rheological response at arbitrary strains and strain rates. 
Despite these interesting future directions for study, our explanation for the corrections to the monolithic Flory behavior induced by the affine facet slip provides a starting point for future quantitative models of the mechanical response of these packings.

\section{Acknowledgements}
We thank Alberto Fernandez-Nieves, Emanuela Zaccarelli, Lorenzo Rovigatti, Michel Cloitre, Robin Selinger, and especially Massimo Pica Ciamarra, Max Bi, and Jeffrey Sokoloff for useful discussions in the early stages of this work.
We thank Manuel Valera for their help with code development and optimization.
We thank MGHPCC for computing support.
This material is based upon work supported by the National Science Foundation under Grant No. CMMI-1822020. 

\bibliography{references_hydrogel_from_proposal,refs-hydrogel-new}
\end{document}